\newtheorem{theorem}{Theorem}
\newtheorem{definition}[theorem]{Definition}
\newtheorem{lemma}[theorem]{Lemma}

\documentclass[12pt]{article}

\begin{document}

\title{A Geometric Model of Information Retrieval Systems}
\author{\textbf{Myung Ho Kim} \\
19 Gage Road, \\
East Brunswick, NJ 08816 \\
mhkim@math1.math.auckland.ac.nz}
\date{Dec 4 1999}
\maketitle

\begin{abstract}
This decade has seen a great deal of progress in the development of
information retrieval systems. Unfortunately, we still lack a systematic
understanding of the behavior of the systems and their relationship with
documents. In this paper we present a completely new approach towards the
understanding of the information retrieval systems. Recently, it has been
observed that retrieval systems in TREC 6 show some remarkable patterns in
retrieving relevant documents. Based on the TREC 6 observations, we
introduce a geometric linear model of information retrieval systems. We then
apply the model to predict the number of relevant documents by the retrieval
systems. The model is also scalable to a much larger data set. Although the
model is developed based on the TREC 6 routing test data, I believe it can
be readily applicable to other information retrieval systems. In Appendix,
we explained a simple and efficient way of making a better system from the
existing systems.
\end{abstract}

\section{Introduction}

Presently, internet is being more and more frequently used for information
retrieval for a wide variety of activities ranging from routine tasks such
as shopping and reading newspaper to esoteric researches. As the internet
was about to become ubiquitous and grow at an explosive rate, in 1992,
National Institute of Standards \& Technology, Information Access \& User
Interfaces Division(IAUI) and Defense Advanced Research Projects Agency
(DARPA), joined to start a conference named ''TREC (Text Retrieval
Conference)'' for exchanging technology and research. One of the most
exciting events in TREC conference is the performance (of retrieving
relevant documents) competition of participating institutes all over the
countries. To get some idea, we need to know the following competition
procedure.\vspace{0.5 mm}

\textbf{Step 1}. To thirty-one participating systems, TREC 6 provided

i) 47 topics

ii) 120,653 documents, which each system was supposed to search through for
picking up relevant documents, for each of 47 topics

\textbf{Step 2}. Thirty-one computer systems submitted 1000 ranked retrieved
documents for each topic (a document of rank 1 is considered as the most
relevant by a system). The total is 31 x 47 x 1000 lists of pairs of topic
and document id.

\textbf{Step 3}. For each topic, TREC collected top 100 documents for each
system and judged to their relevance to the given topic. For example, for
topic 1, if F128 is retrieved by system att97re as 20th document, it judged
whether the document for topic 1 was relevant. The performance results of 31
systems were announced. This is the data I worked on and analyzed.

For some topics, most of systems did not do well in retrieving relevant
documents. As a matter of fact, there were a few relevant documents to be
retrieved and it is natural for systems to choose sufficient number of
relevant documents for determining a certain pattern of behavior. In other
words, the collection of documents is not suitable for some topics. For
testing the model, this is the reason why we chose the best five systems and
six topics.( See Section 4).

\section{A Geometric Linear Model}

For a set of retrieval systems and a set of relevant documents for a topic,
we describe a geometric approximation model. Before we go into abstract
setup, it is good to see the overall picture: For example, suppose that
there are three systems, A, B, C and a topic, ''car''. And suppose that
there is a collection of 1000 documents. Let those A, B, and C retrieve 100
documents, about car, from the collection. Then by the lemmas below, we can
calculate all the angles, in this case, three angles and three ratios (a
kind of relative detection power for each pair of systems). Now, if we have
the same three systems and a collection of one million documents (Remark:
This collection has nothing to do with the collection of one thousand
documents), choose any one of those A, B, C system and let it retrieve
relevant documents from those million documents. Let the number of relevant
documents retrieved be n. Then we may estimate all the numbers of relevant
documents retrieved by the remaining two systems, by multiplying ratios
calculated in the previous steps. Moreover, we may estimate the total number
of relevant documents retrieved by three systems.

Now, we are ready for the model. Let $R^n$ be the Euclidean n-dimensional
space with the usual inner product, i.e.,

\[
v\cdot w=\sum_{i=1}^nv_iw_i 
\]

where $v,w\in R^n$ and $n$ is an integer.

\textbf{Assumption 1}. For each topic, in $R^n$ each retrieval system is
represented by a line passing through the origin and a set of all lines(i.e.
systems) are linearly independent.

\textbf{Assumption 2.} The set of relevant documents is represented as a
vector, which we call the \emph{relevant set vector} . Moreover, consider
the projection of the relevant set vector to the line represented by a
system. We denote it by $Pr(v),$where $v$ is the \emph{relevant set vector}
and

\textbf{Assumption 3.} Assume its length (the absolute value of $Pr(v)$, $|$%
Pr(v)$|$) as an approximation of the number of relevant documents retrieved
by the system. Before we make the fourth assumption in section 3, we need to
mention the following useful lemma. By using it, given the angle between a
pair of projection vectors, we may estimate the number of relevant documents
retrieved by the pair.

\begin{lemma}
Suppose that there are a vector $u$ in a plane and two straight lines
through the origin as below. Let two vectors $v$ and $w$ be the projections
of $u$ to those two lines and $\theta $ be the angle between those vectors $v
$ and $w$ . Then $u$ may be expressed as follows

\begin{center}
\begin{picture}(200, 200)(0,0)
\put(0, 200){i)Case I }
\put(50, 0){\vector(2,1){120}}

\put(20, 10){\line(1, 2){10}}
\put(14, 32){\line(3, -1){17}}

\put(50, -15){$O$}
\put(20, 0){$v$}
\put(110, 20){$w$}
\put(75, 70){$u$}
\put(60, 10){$\theta$}
\put(50, 0){\line(-2, -1){40}}
\put(50, 0){\vector(1,4){50}}
\put(50, 0){\line(3, -1){90}}
\put(50, 0){\vector(-3, 1){43}}
\line(1,2){105}
\put(70, 50){\line(-1, 2){75}}
\put(50, 70){\line(2, 1){10}}
\put(50, 70){\line(1,-2){7}}

\end{picture}
\end{center}

\vspace{0.5 in}

\[
u=\frac{\sqrt{|u|^2-|v|^2}}{\sin \theta }\frac w{|w|}+\frac{\sqrt{|u|^2-|w|^2%
}}{\sin \theta }\frac v{|v|}
\]

\begin{center}
\begin{picture}(200, 190)(0,0)
\put(5, 150){ii)Case II}

\put(100, 0){\vector(4,1){133}}
\put(100, 0){\line(4,1){138}}

\put(100, 0){\line(-4, -1){50}}
\put(100, 0){\vector(1,1){100}}
\put(143, -13){\line(1, 2){56}}
\put(138, -11){\line(1, 2){5}}
\put(140, -1){\line(4, -1){10}}
\put(100, 0){\line(4, -1){60}}
\put(100, 0){\vector(4, -1){46}}
\put(100, 0){\line(-4, 1){50}}
\put(233, 33){\line(-1, 2){33}}
\put(225, 31){\line(-1, 2){7}}
\put(227, 45){\line(-4, -1){10}}
\put(98, -13){$O$}
\put(120, -20){$w$}
\put(140, 55){$u$}
\put(120, -4){$\theta$}
\put(190, 10){$v$}
\end{picture}
\end{center}

\vspace{0.5 in} 
\[
\ u=\frac{\sqrt{|u|^2-|v|^2}}{\sin \theta }\frac w{|w|}-\frac{\sqrt{%
|u|^2-|w|^2}}{\sin \theta }\frac v{|v|}
\]

\vspace{0.5 mm}

\begin{center}
\begin{picture}(200, 200)(0,0)
\put(5, 150){iii)Case III}

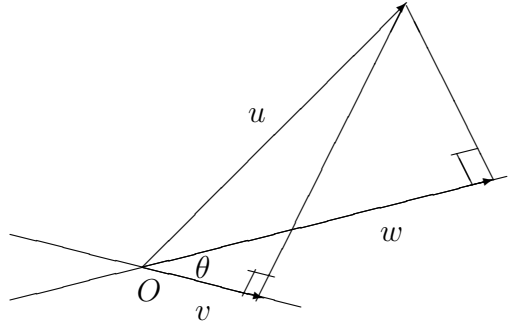
\put(90, 0){\vector(4,1){133}}
\put(90, 0){\line(4,1){138}}

\put(90, 0){\line(-4, -1){50}}
\put(90, 0){\vector(1,1){100}}
\put(133, -13){\line(1, 2){56}}
\put(128, -11){\line(1, 2){5}}
\put(130, -1){\line(4, -1){10}}
\put(90, 0){\line(4, -1){60}}
\put(90, 0){\vector(4, -1){46}}
\put(90, 0){\line(-4, 1){50}}
\put(223, 33){\line(-1, 2){33}}
\put(215, 31){\line(-1, 2){6}}
\put(217, 45){\line(-4, -1){10}}
\put(88, -13){$O$}
\put(110, -20){$v$}
\put(130, 55){$u$}
\put(110, -4){$\theta$}
\put(180, 10){$w$}
\end{picture}
\end{center}

\vspace{0.5 in} 
\[
u=-\frac{\sqrt{|u|^2-|v|^2}}{\sin \theta }\frac w{|w|}+\frac{\sqrt{%
|u|^2-|w|^2}}{\sin \theta }\frac v{|v|}
\]

\vspace{0.5 mm}

and

\[
|u|=\frac{\sqrt{|v|^2+|w|^2-2|v||w|\cos \theta }}{\sin \theta }
\]
$.$
\end{lemma}

\vspace{0.5 mm}

\textbf{Proof)} Refer to any calculus book.

\section{Behavior of the systems in TREC 6}

According to TREC 6 data, for each topic, it is evident that all systems
showed surprising patterns in retrieving relevant documents (\cite{KK:test}
and \cite{TT:tr}). Here is one of such patterns, which motivated us to write
this paper and lead to the last and most important assumption.

Notations : For topic $t$, let $a_1\left( r,t\right) $ , $a_2\left(
r,t\right) $ and $a_{12}\left( r,t\right) $ be the accumulated numbers of
relevant documents retrieved by system1, system2, and by both, up to certain
stage $r$ (In our setting, $r$ represents rank.) More precisely, if we have
system1 and system2 retrieved relevant documents, for a topic $t$, as
follows:

\begin{center}
\begin{tabular}{lllll}
\textbf{r} & \textbf{system1} &  & \textbf{system2} &  \\ 
1 & F234 & 1 & F345 & 1 \\ 
2 & F345 & 1 & F1789 & 0 \\ 
3 & F4 & 1 & F56 & 1 \\ 
4 & F78 & 1 & F3590 & 0 \\ 
5 & F1789 & 0 & F23 & 1 \\ 
6 & F23 & 1 & F4 & 1 \\ 
7 & F57 & 0 & F983 & 0
\end{tabular}
\end{center}

where in the third and fifth columns, 1 means ''relevant'' and 0 means
''irrelevant'', then

$a_1\left( 1,t\right) $ = 1, $a_1\left( 2,t\right) $=2, $a_1\left(
3,t\right) $=3, $a_1\left( 7,t\right) $=5, $a_2\left( 1,t\right) $ =1, $%
a_2\left( 2,t\right) $= 1.....$a_2\left( 7,t\right) $= 4, and $a_{12}\left(
1,t\right) $=0, $a_{12}\left( 2,t\right) $=1.......,$a_{12}\left( 7,t\right) 
$=3.

Remarkably we have found that the ratio of $a_1\left( r,t\right) $ to $%
a_2\left( r,t\right) $ is almost constant over all $r$ and, even so is the
ratio of $a_{12}\left( r,t\right) $ to $a_2\left( r,t\right) $. For example,
for the topic $62$, if we run a linear regression for the numbers of
relevant documents retrieved by two systems -\emph{Cor6R1cc} and \emph{ETH6R2%
}- up to rank (or $r$) $100$, the graph has the slope $0.95(=$ $a_1/a_2$)
and $0.58(=a_{12}/a_2)$ with high accuracy, i.e. rsquare 0.9994 and 0.9910
respectively. This implies that the ratio, so called '\emph{relative
detection powers}' of those two systems, is almost constant. The results are
similar for all topics. From this remarkable observed facts, we assume the
most interesting assumption.

\textbf{Assumption 4}. For each topic, all the ratios of $a_1\left(
r,t\right) $ to $a_2\left( r,t\right) $ and $a_{12}\left( r,t\right) $ to $%
a_2\left( r,t\right) $ are constant over all r.

\section{Application}

In this section, we apply the model (which consists of 4 \textbf{assumptions}%
) to calculate the total number of relevant documents retrieved by several
systems. To get an idea, it is enough to consider three systems, since we
may reduce any cases to the case of two systems.

Suppose that there are three systems(i.e. lines) $s_1,$ $s_2,$ $s_3$ their
relative detection powers(i.e. ratios of $a_1\left( r,t\right) $ to $%
a_2\left( r,t\right) $ and $a_{12}\left( r,t\right) $ to $a_2\left(
r,t\right) $) are known for each topic. Suppose that a relevant set vector
lies in the space spanned by the three systems. Then, given the number of
relevant documents retrieved by any one of the three systems for some topic,
by applying the model above, we may estimate the total number of relevant
documents retrieved by the three systems for the given topic, as follows. To
do it, first we need to calculate $cosine$ value of the angle between the
lines represented by systems(More precisely, the angle between the
projection vectors of the given \emph{relevant set vector} to the lines).

\begin{lemma}
Let $a_1,$ $a_2$ and $a_{12}$ be the numbers of relevant documents retrieved
by system1, system2 and both of two systems. Let $\theta $ be the angle
between two systems. And let $k$ and $\rho $ be the ratios $\frac{a_1}{a_2}$
and $\frac{a_{12}}{a_2}$ respectively. Then $a_1=ka_2$ $a_{12}=\rho a_2$,and
we have

\vspace{0.5 mm}

i)Case I

\vspace{0.5 mm}

\[
\cos \theta =\frac{a_1a_2-\sqrt{\left( a_1+a_2-a_{12}\right) ^2-a_1^2}\sqrt{%
\left( a_1+a_2-a_{12}\right) ^2-a_2^2}}{\left( a_1+a_2-a_{12}\right) ^2}
\]

\ \vspace{0.5 mm}

\begin{center}
or
\end{center}

\ \vspace{0.5 mm}

\[
=\frac{k-\sqrt{\left( k+1-\rho \right) ^2-k^2}\sqrt{\left( k+1-\rho \right)
^2-1}}{\left( k+1-\rho \right) ^2}
\]

ii)Case II and III

\vspace{0.5 mm}

\[
\cos \theta =\frac{a_1a_2+\sqrt{\left( a_1+a_2-a_{12}\right) ^2-a_1^2}\sqrt{%
\left( a_1+a_2-a_{12}\right) ^2-a_2^2}}{\left( a_1+a_2-a_{12}\right) ^2}
\]

\emph{\vspace{0.5 mm} }

or

\emph{\vspace{0.5 mm} }

\[
=\frac{k+\sqrt{\left( k+1-\rho \right) ^2-k^2}\sqrt{\left( k+1-\rho \right)
^2-1}}{\left( k+1-\rho \right) ^2}
\]
\vspace{0.5 mm}
\end{lemma}

\textbf{Proof)} It follows from the {\emph{cosine}} sum formula to two
triangles in the figures of \textbf{lemma 1}.

\vspace{0.5 mm}

Suppose $v_i$ is the projection vectors of systems and $n_i$ is the number
of relevant documents retrieved by system s$_i$ for $i=1,2,3$, in other
words, $|v_i|=n_i$.As we mentioned in the \textbf{section 3, behavior of the
systems in TREC 6}, we can obtain all the $k$ and $\rho $ values for all
pairs of systems for each topic with the help of a statistical software such
as SAS. Suppose only $n_1$ is known. Then $n_2$ and $n_3$ are estimated as $%
n_1$times proper $k$. By the \textbf{lemma} \textbf{1}, we get the \emph{%
relevant sum vector } $u$ (in the plane generated by $v_1$and $v_2)$ of two
projection vectors $v_1$, $v_2$. Again by applying \textbf{lemma 1 }to $u$
and $v_3$, we get the \emph{relevant sum vector } $w$(in the three
dimensional space generated by $v_1,v_2,v_3)$ for the $v_1,v_2,v_3$ and the
absolute value of the sum vector $w$ will be the total number of relevant
documents retrieved by the three systems. In this way, this procedure can be
applied to any finite number of systems.\newline

To test the geometric model, we chose the best five(Cor6R1cc, ETH6R2,
att97re city6r1 pirc7R2) out of 31 systems in TREC 6, and seven topics (189,
161, 111, 10002, 62, 54, 154) of most relevant documents. By assuming \emph{%
Case I} for all pairs, the following results are obtained.

\vspace{0.5 mm}

\begin{center}
\textbf{Estimation by the Geometric Model} \vspace{2 mm} $
\begin{tabular}{llllll}
\textbf{Topic} & \textbf{One} & \textbf{Two} & \textbf{Three} & \textbf{Four}
& \textbf{Five} \\ 
54 & 89 & 100.643 & 106.624 & 110.110 & 125.084 \\ 
62 & 81 & 116.067 & 123.336 & 181.108 & 312.631 \\ 
111 & 74 & 137.786 & 160.335 & 194.045 & 263.126 \\ 
154 & 83 & 110.412 & 113.516 & 121.442 & 124.426 \\ 
161 & 71 & 78.726 & 87.896 & 102.343 & 110.606 \\ 
189 & 90 & 150.801 & 180.346 & 247.611 & 286.277 \\ 
10002 & 69 & 102.358 & 115.578 & 119.521 & 120.998
\end{tabular}
$
\end{center}

\pagebreak

\begin{center}
\textbf{Numbers of True Relevant Documents }\nopagebreak
\vspace{2 mm}
$%
\begin{tabular}{llllll}
\textbf{Topic} & \textbf{One} & \textbf{Two} & \textbf{Three} & \textbf{Four}
& \textbf{Five} \\ 
54 & 89 & 101 & 111 & 113 & 114 \\ 
62 & 81 & 116 & 129 & 170 & 217 \\ 
111 & 74 & 141 & 162 & 180 & 208 \\ 
154 & 83 & 113 & 117 & 128 & 130 \\ 
161 & 71 & 80 & 91 & 102 & 102 \\ 
189 & 90 & 152 & 174 & 200 & 216 \\ 
10002 & 69 & 103 & 114 & 121 & 126
\end{tabular}
$
\end{center}

\emph{\ \vspace{2 mm} }

Here \emph{One} means the total number of documents by Cor6R1cc, \emph{Two}
by Cor6R1cc and ETH6R2, \emph{Three} by Cor6R1cc ETH6R2 and att97re, \emph{%
Four} by Cor6R1cc ETH6R2 att97re and city6r1, \emph{Five} by Cor6R1cc ETH6R2
att97re city6r1 and pirc7R2.

\section{Conclusion}

As the table in \textbf{section 4} showed, the geometric model works
well for three-systems, for all topics (Note that, since we get the angle
from the number of relevant documents retrieved by two systems, the model is
expected to fit well with two systems). To make this geometric model work
better for all topics and more than three systems, further investigation is
required to find the best \emph{combination of cases} (See cases in 
\textbf{lemma 1} and recall that in our experiments, we assume \emph{%
case 1} for all). Note that, to each system, by associating a line in an
Euclidean space, a new concept of ''independence'' of systems was
introduced. In other words, if the set of representing lines is linearly
independent, then we might say that the corresponding systems are
independent. It seems that there is a subtle point in this concept, because
all systems have a tendency to resemble each other. In \textbf{Appendix}%
, we present a very simple and interesting scoring method, in other words,
describe a way of making a new system from the existing systems. The
experimental results say that this method, i.e., new system, performs better
than each individual system, the principle of which is widely applicable.

\section*{\textbf{\ Acknowledgements}}

While I am visiting the department of statistics at Rutgers University from
Oct. 1998 to Sep. 1999, I would like to thank for their hospitality and
giving me the chance to present a talk for this paper. I am indebted to
Vladimir Menkov, Gene Kim and W. Shim about technical supports and
criticism. I also give special thanks to NIST for making TREC 6 results
available. I want to express my appreciation for the thoughtful advice and
suggestions by the reviewer.

\appendix

\section{On a Scoring Method of Retrieved Documents by Systems}

We introduce a new scoring algorithms based on the ranks given by systems
for better results over all individual systems. Since each system has its
own scoring method and it needs to be normalized in some way. The ranks
given by systems are our natural choice and here is our definition of the
new scoring method.

\subsection{Definition of scoring and results}

\begin{definition}
Given a pair of topic and document, first, we interpret 100-rank (or
101-rank) as its score given by each system, if ranks of a pair are less
than or equal to 100, otherwise 0. Second, compute the average of the scores
and assign it to the score of the document for a given topic. By the
definition above, we re-scored all pairs (document id, rank) judged by TREC
6 and chose top100 documents for each topic.
\end{definition}

Here is the table by our new scoring method.

T : topic

$f_{31}$: the system of combining thirty-one, i.e., by averaging all scores
from 31 systems

$B_{31}$ : number of systems which perform better than $f_{31}$ among 31
systems

$f_6$ : the system of combining the six best systems, i.e., by averaging the
scores from only the best six systems

$B_6$ :number of systems which perform better than $f_6$

$G$ :number of relevant documents retrieved by the whole 31 systems

\begin{tabular}{llllll}
\textbf{T} & \textbf{$f_{31}$} & \textbf{$B_{31}$} & \textbf{$f_6$} & 
\textbf{$B_6$} & \textbf{$G$} \\ 
1 & 39 & 0 & 35 & 4 & 51 \\ 
3 & 45 & 3 & 48 & 0 & 70 \\ 
4 & 41 & 4 & 45 & 3 & 70 \\ 
5 & 6 & 0 & 6 & 0 & 7 \\ 
6 & 54 & 5 & 55 & 4 & 146 \\ 
11 & 48 & 7 & 45 & 7 & 150 \\ 
12 & 78 & 6 & 81 & 2 & 270 \\ 
23 & 4 & 3 & 4 & 3 & 6 \\ 
24 & 11 & 14 & 17 & 3 & 34 \\ 
44 & 4 & 0 & 4 & 0 & 4 \\ 
54 & 94 & 1 & 92 & 3 & 164 \\ 
58 & 16 & 2 & 17 & 1 & 18 \\ 
62 & 86 & 6 & 85 & 7 & 368 \\ 
77 & 13 & 1 & 13 & 1 & 16 \\ 
78 & 40 & 1 & 39 & 1 & 45 \\ 
82 & 61 & 0 & 59 & 2 & 80 \\ 
94 & 48 & 3 & 49 & 3 & 174 \\ 
95 & 36 & 5 & 38 & 1 & 123 \\ 
100 & 80 & 4 & 82 & 4 & 179 \\ 
108 & 71 & 5 & 77 & 2 & 293 \\ 
111 & 90 & 5 & 91 & 1 & 480 \\ 
114 & 37 & 1 & 38 & 1 & 57 \\ 
118 & 38 & 0 & 36 & 2 & 83 \\ 
119 & 20 & 5 & 22 & 2 & 76 \\ 
123 & 44 & 2 & 43 & 2 & 62 \\ 
125 & 18 & 3 & 23 & 0 & 27 \\ 
126 & 15 & 1 & 15 & 1 & 19 \\ 
128 & 65 & 2 & 70 & 1 & 281 \\ 
142 & 57 & 1 & 51 & 3 & 200 \\ 
148 & 79 & 11 & 86 & 11 & 241 \\ 
154 & 88 & 6 & 92 & 0 & 168 \\ 
161 & 84 & 6 & 88 & 2 & 118 \\ 
173 & 11 & 3 & 15 & 0 & 16 \\ 
180 & 2 & 9 & 5 & 5 & 17 \\ 
185 & 16 & 2 & 16 & 2 & 18\\
187 & 10 & 7 & 10 & 7 & 19 \\ 
189 & 91 & 4 & 90 & 5 & 667 \\ 
192 & 7 & 0 & 4 & 13 & 7 \\ 
194 & 3 & 0 & 3 & 0 & 4 \\ 
202 & 87 & 5 & 88 & 4 & 534 \\ 
228 & 29 & 4 & 28 & 5 & 58 \\ 
240 & 29 & 0 & 25 & 2 & 121 \\ 
282 & 16 & 0 & 17 & 0 & 28 \\ 
10001 & 70 & 1 & 66 & 2 & 127 \\ 
10002 & 73 & 3 & 72 & 3 & 267 \\ 
10003 & 47 & 3 & 55 & 0 & 72 \\ 
10004 & 12 & 4 & 14 & 0 & 17
\end{tabular}

\section{ Implications}

First, information retrieval systems are more or less independent, which
means they retrieve documents in their own way. Second, more importantly,
systems tend to choose relevant documents more than irrelevant ones, i.e.,
give higher scores to relevant ones (standard deviation are smaller). This
observation has an intuitive appeal if we believe that noise, physical or
informational, tend to be uncorrelated. This is believed to be the reason
that $f_{31}$ and $f_6$ systems do better than each individual system.

\end{document}